\documentstyle[12pt]{article}

\begin{document}

\baselineskip=7mm

\begin{center}

{\bf Angular momentum $I$ ground state probabilities of  boson systems
interacting by random interactions}

\vspace{0.2in}

{Y. M. Zhao$^{a,b}$,  A. Arima$^{c}$, and N. Yoshinaga$^a$,  }

\vspace{0.2in}
{
$^a$ Department of Physics, Saitama University, Saitama-shi, Saitama 338 Japan \\
$^b$ Department of Physics,  Southeast University, Nanjing 210018 China \\
$^c$ The House of Councilors, 2-1-1 Nagatacho, 
Chiyodaku, Tokyo 100-8962, Japan }

\end{center}

In this paper we  report our systematic calculations of 
angular momentum $I$ ground state probabilities ($P(I)$) 
of  boson systems with  spin $l$ 
in the presence of random two-body interactions.
It is found that
the $P(0)$ dominance is usually not 
true for a system with an odd number of bosons, while 
it is valid for an even number of bosons, which indicates 
that the $P(0)$ dominance is partly connected to the 
even number of identical particles. It is also noticed that 
the $P(I_{max})$'s of bosons with  spin $l$  do not 
follow  the $1/N$ ($N=l+1$, referring to the number 
of independent two-body matrix elements) relation.  The  
properties of the $P(I)$'s obtained in boson systems 
with  spin $l$ are   discussed.
          

\vspace{0.4in}
        
\newpage

It was  discovered  a few years ago \cite{Johnson} that  
 the dominance of the $0^+$ ground state (0 g.s.) of even 
fermion systems is obtained by diagonalizing a scalar two-body Hamiltonian 
which is randomly determined.  
Many works have been done in the context of this discovery [2-16]. 
For example, many efforts were
made concerning  the origin of this observation  and
 generic properties  
of many-body systems in the presence of random interactions.

In the case of  $d$-boson systems 
the  0 g.s. probability (denoted as $P(0)$) 
was found to be close to $0\%$ periodically when  the 
number of $d$ bosons $n=6k \pm 1$, where $k$ is a 
positive integer \cite{Zhaox1}.
This finding of systematic counter
examples of 0 g.s. dominance in
$d$-boson systems suggests that 
the breaking of the  0 g.s. dominance
might be related to   certain  features associated
with bosonic degrees of freedom and thus 
could be easily seen  also for bosons with other $l$.
To see whether this is true or not we report 
  in this paper  systematic calculations 
of boson systems with    spin $l$ in the presence of random 
interactions.

The Hamiltonian that we take  is as follows:
\begin{equation}
  H_d =  \sum_{l}  
  \frac{1}{2} \sqrt{2l+1} G_L \left( \left(b_l^{\dagger} \times
  b_l^{\dagger}  \right)^{(L)} \times \left( \tilde{b}_l \times \tilde{b}_l
  \right)^{(L)} \right)^{(0)}    \label{hamiltonian-1} 
\end{equation}
where $b_l^{\dagger}$ is the creation operator of a boson with  spin $l$, and
$\tilde{b}_l$ is the time-reversal  of the annihilation operator $b_l$.
The $G_L$'s are two-body interactions between the bosons, and 
are assumed to follow a two-body random ensemble (TBRE), i.e., 
the $G_L$'s are a set of random numbers with a distribution function
\begin{equation}
  \rho(G_L) = \frac{1}{\sqrt{2\pi}} {\rm exp}(-G_L^2/2),~~~L=0,2, \cdots, 2l.
    \label{tbre}
\end{equation}
All the results in this paper are obtained by using 1000 sets of the above
two-body random interactions.

We first look at the case  of four bosons with  spin $l$.
These systems are interesting and 
worthwhile to study  separately, 
because their  fermionic counterparts, namely, 
four fermions in a single-$j$ shell,
 were studied extensively
 \cite{zelevinsky,Zhaox1,Zhaox0} since these 
 systems are the simplest and non-trivial ones to investigate 
the 0 g.s. dominance  of fermions. 
Fig. 1 shows a few important angular momentum $I$
ground state probabilities  versus $l$ for systems
with boson number $n$=4. 
One sees that  the pattern of $I$ g.s. probabilities ($P(I)$'s) of 
 four bosons with  spin $l$ is very similar 
to that of four fermions in a single-$j$ shell. 
For instance,   $P(0)$'s vs. $l$ stagger with an interval of 
$\delta_l = 3$; $P(0)$'s are dominant over other $P(I)$'s except for only 
two  cases: $l=2$ and 8; $P(I_{max})$'s decrease with $l$.  A
new feature is that the $P(l)$'s 
are  considerably large (small) when $l$ is even (odd), i.e., the 
$P(l)$'s exhibit an even-odd staggering behavior.
We should  be aware that $I=l$ is possible in the four boson case, 
while in the case of four 
fermions there are no $I$=$j$ states.

The 0 g.s. dominance in the case of  four bosons with 
 spin $l$ can be understood based on the following observation: 
 there is {\it only} one non-zero eigenvalue (denoted as $E_0^{L(l)}$)
 among all eigenvalues 
 of $I=0$ states  if  only one of $G_L = -1$ and others are zero. This 
 observation was proved recently 
 by constructing the $I=0$ states using a pair basis \cite{prc7}. 
This means that there is a very large probability for
$I=0$ states to give the lowest eigenvalue, compared to  
other $I$ states for which 
there are many non-zero eigenvalues. According to our 
empirical rule \cite{Zhaox1},
$P(I) = {\cal N}_I/N$, where ${\cal N}_I$ is the number of  
times that the ground state has angular momentum $I$ when one of the
two-body matrix elements is $-1$ and others are zero, and $N$ is the number
of two-body matrix elements. The  $P(0)$'s are therefore expected to be
larger than all the other $P(I)$'s (with only two exceptions of $l=2$
and 8).

In Fig. 1 one sees that the increase of $P(0)$'s ``coincides" with that 
of $D_0^{(l)}$, the  number of $I=0$ states of four bosons 
with  spin $l$. $D_0^{(l)}$ is 1,  1, 1, 2, 2, 2, 3, 3, 3,
4, 4, 4, $\cdots$ for $l$=0, 1, 2, 3, $\cdots$, etc. 
Although we still   can not  prove  this correlation, 
below we give an argument based on our  empirical rule \cite{Zhaox1}.
From the sum  rule of diagonal matrix elements \cite{Yoshinaga},
one obtains that $\sum_L E_0^{L(l)} = -\frac{1}{2} n(n-1) D_0^{(l)}
= -6 D_0^{(l)}$,
where  $n$ is the number of 
bosons in the system and $n=4$ here. 
This relation means that the sum of $E_0^{L(l)} \times (-1)$
increases with $D_0^{(l)}$,  suggesting 
that ${\cal N}_0$ increases  with 
$D_0^{(l)}$ simultaneously. Because $P(0) = {\cal N}_0 / N$, 
a regular increase of $D_0^{(l)}$
of four bosons with  spin $l$ \cite{Ginocchio}   produces
a regular staggering of  the $P(0)$'s. It is noted that
a similar ``coincidence" of the   $P(0)$'s staggering 
 with increasing number of $I$=0 states  was firstly observed 
in \cite{zelevinsky,Zhaox1} for four fermions in a 
single-$j$ shell, but without an explanation or argument.

We next go to systems with arbitrary  particle number $n$, 
in order to investigate  general features of the $P(I)$'s 
of boson systems with  spin $l$ in the presence of random interactions. 
Here a  particularly interesting question
should  be asked concerning the $P(I)$'s for an odd number of particles:
 which angular momentum $I$  g.s.  dominance will appear, $I=0$  or $I=l$?
 In systems with an odd number of
fermions in a single-$j$ shell, there is no $I$=0 state. One 
expects  $I=j$ g.s. probabilities  to 
be   large, because one easily relates  large $j$ g.s. probabilities 
to large 0 g.s. probabilities for the neighboring  
even $n$, and the calculations using different 
 types of random interactions showed \cite{zelevinsky,Zhaox1} 
 that this is indeed the case.  For systems with odd $n$,
 however, one may have $I$=0 states and thus
 it is unknown  a priori whether the $I=0$ or $I=l$ g.s. dominance  occurs
 in these systems.

Fig. 2 shows  the $P(l)$'s and  $P(0)$'s in 
boson systems with $l=$4 and  6, and 
$n$ running as large as possible.  
For odd $n$ and $l=1$ or $n=3$ and any odd $l$, 
 there is no $I=0$ state;
for odd $n$ and $l=3$, 5, 7 and 9, the $I=0$ states do not exist
unless $n\ge 15$, 
9, 7, and 5, respectively. For fifteen bosons with $l=3$, 
nine bosons with $l=5$, seven bosons with $l=7$ and 
five bosons with $l=9$ or 11,  $P(0) \sim 0 \%$ always 
according to our calculations using 1000 sets of a TBRE Hamiltonian. 
From Fig. 2 and these odd-$l$ cases we 
conclude that the $P(0)$'s are in general 
much less than the corresponding $P(l)$'s 
when $n$ is   odd. On the other hand, the $P(0)$'s are
mostly larger than the  $P(l)$'s when $n$ is  even. 
   This result  indicates that
the 0 g.s. dominance is robust
for systems with  even $n$, but  not true generally
if $n$ is   odd. In the latter case,  we observe that 
the 0 g.s. dominance is  easily lost, with a few exceptions. 
We therefore expect that the 0 g.s. dominance 
is partly connected to the even number of particles. 

It was shown in previous calculations that the 
0 g.s. dominance  may appear in  boson systems with odd $n$, 
such as $sp$ \cite{Kus,Bijkerx}, $sd$ \cite{Bijker,Bijkerx,Zhaox1},
and $sdg$ \cite{Zhaox1} boson systems. However, 
to a large extent,  the very large
$P(0)$ therein are associated with  the $s$ boson condensation,  which  
contributes around $40\%$ of $P(0)$  \cite{Bijkerx}. 
In other words, $sp$ and $sd$ systems
are very special systems in which the $s$ boson condensation 
produces 0 g.s. dominance  when $n$ is odd.  
Without $s$ bosons, the $P(0)$'s of those systems with odd $n$ would be  
drastically smaller than those with $s$ bosons,  and g.s. probabilities
for other $I$'s, such as $I=l$ or $I=I_{max}$,  would 
be much larger. 
 It would be very interesting to carry out 
systematic calculations of $P(I)$'s for both even and odd numbers  
of bosons  with many $l$'s.

Now we come to the $I_{max}$ g.s. probabilities
of bosons with  spin $l$ interacting by a TBRE Hamiltonian. 
It was shown in \cite{Zhaox1} that the $P(I_{max})$'s are associated with the
number of independent two-body matrix elements, and not
sensitive to the particle numbers of the system. 
For fermions in a single-$j$ shell, $P(I_{max}) \sim 1/N$, where
$N=j+\frac{1}{2}$.
This relation is also applicable to $sd$ and $sdg$ boson systems. 
Thus it is interesting to check whether the 
$P(I_{max})$'s of more general boson systems  vs. $l$ show a similar 
pattern or not. 

Fig. 3 shows the $P(I_{max})$'s vs. $l$ with $n$ ranging from three to six.
One sees large deviations of  the $P(I_{max})$'s
calculated by diagonalizing a TBRE Hamiltonian from those 
predicted by the $1/N=1/(l+1)$ relation. 
  The  $P(I_{max})$'s are systematically larger  
than $1/N$ (when  spin $l$  is small
the agreement is rather good, however), and increase
with $n$.

An argument why the  behavior of the $P(I_{max})$'s is different for 
bosons and fermions is as follows.
It is known from previous studies \cite{Zhaox1}
that the large $P(I_{max})$ 
comes from a gap produced by the pairing
interaction  $G_{L_{max}}$ for bosons with  spin $l$, 
and the pairing interaction $G_{J_{max}}$  for fermions in a single-$j$ shell, where
$L_{max}=2l$ and $J_{max}=2j-1$, respectively. 
For both  cases   one may evaluate 
the gap associated with $G_{J_{max}}$ or $G_{L_{max}}$ 
using  analytical formulas of $(E_{I_{max}-2} -  E_{I_{max}})$,
because here the state with $I_{max}$ ($I_{max}-2$)
is found to be the ground (first
excited) state if   $G_{J_{max}}$ or $G_{L_{max}}$ is $-1$. 
We obtain $(E_{I_{max}-2} -  E_{I_{max}})$ as follows: 
\begin{eqnarray}
  {\rm boson ~~ systems}: &&
\frac{2ln-1}{4l-1};
\nonumber  \\
  {\rm fermion ~~ systems, } ~ n=4: &&
\frac{3}{8} + \frac{105}{108(4j-7)} + \frac{135}{64(4j-5)}
\nonumber  \\
&& 
+ \frac{63}{128(4j-3)};
\nonumber  \\
  {\rm fermion ~~systems, } ~ n=5: &&
\frac{35}{128} + \frac{2205}{2048(4j-9)}
+ \frac{5145}{2048(4j-7)}
\nonumber  \\
&& 
- \frac{1785}{2048(4j-5)}
-  \frac{189}{2048(4j-3)};
\nonumber  \\
  {\rm fermion~~ systems, } ~ n=6: &&
\frac{27}{128} + \frac{10395}{8192(4j-11)} + \frac{2835}{1024(4j-9)}
\nonumber  \\
&& 
- \frac{4725}{4096(4j-7)} - \frac{45}{256(4j-5)}
-  \frac{297}{8192(4j-3)}.
\nonumber  \\
\label{final}
\end{eqnarray}
One easily sees that the gap for  bosons
with  spin $l$   increases  regularly with $n$ 
at an interval$\sim 1/2$  if $l$ is large, while that for  
fermions in a single-$j$ shell is much smaller
(almost one order) in magnitude 
and comparable for different $n$ and $j$.
For instance,  the gap is 0.47, 0.39,  0.35  for
$n=$4, 5 and 6 fermions in a $j=15/2$ shell, respectively, while 
the gap is 2.03, 2.56, 3.07
for $n=4$, 5, 6 bosons with  spin $l=7$, respectively.
According to the empirical rule 
\cite{Zhaox1}, a relatively larger gap makes  the corresponding 
$P(I)$  larger: the larger the gap is,
the larger the corresponding $P(I)$ is.
Therefore, the $P(I_{max})$'s   
of bosons with  spin $l$ can be expected to be larger 
than $1/N=1/(l+1)$, which works well for fermions  and
some boson systems with small $l$'s  
studied in \cite{Zhaox1}, such as $d$, $sd$  and $sdg$ bosons.

To summarize,   in this paper we reported systematic calculations 
of angular momentum $I$ ground state probabilities for   systems of bosons 
with  spin $l$ in the presence of random interactions.  First, 
we found that the 0 g.s. probabilities $P(0)$'s of four bosons with 
spin $l$ exhibit a   pattern  similar to that of 
four fermions in a single-$j$ shell. We explained the 0 g.s. dominance and
the staggering of the $P(0)$'s  based on 
a special property of $I=0$ states when only one of the $G_L$'s are $-1$
and the other two-body matrix elements are zero.

Second, we found that the 0 g.s. dominance is usually not true for  
systems with an odd number of bosons with  spin $l$ 
(though the $I=0$ states may exist), which 
indicates that the 0 g.s. dominance is 
partly connected to the even number of particles.

Third, we found that the $I_{max}$ g.s.
probabilities of bosons with  spin $l$ are systematically larger than 
those predicted by the $1/N$ relation ($N=l+1$ for bosons with  spin $l$), and 
that $P(I_{max})$ increases with boson number $n$. We presented an argument
for this difference between bosons and fermions 
in terms of   a large gap
for bosons systems with $G_{2l}$ being $-1$ and others zero.

One of the authors (ZHAO) would like to thank 
 the Japan Society for the Promotion of Science (ID: P01021)
 for supporting the present work.

------------------------

\vspace{0.3in}

Fig. 1 ~~ The $I$ g.s. probabilities vs. $l$ of four bosons with  spin $l$.
The results are obtained by 1000 runs of a TBRE Hamiltonian, except that
we present the $P(0)$'s (open squares) predicted
by a prescription given in Ref. [10], showing that the
approach of Ref.[10] is reasonably applicable to bosons with  spin $l$.

\vspace{0.4in}

Fig. 2 ~~ The $I=0$ and $I=l$ g.s. probabilities 
versus $n$ of $l=4 $ and 6. 
One sees that $P(0)$ is in general smaller than the 
corresponding $P(l)$ when $n$ is an odd number, indicating that
the 0 g.s. dominance is partly associated with the even number
of particles. 

\vspace{0.4in}

Fig. 3. ~~ The  $P(I_{max})$'s vs. $l$ with $n$
ranging from three to six. It is seen that
the $P(I_{max})$'s are systematically larger than
$1/(l+1)$, and increase with $n$ but seem to saturate 
around $n=6$. 

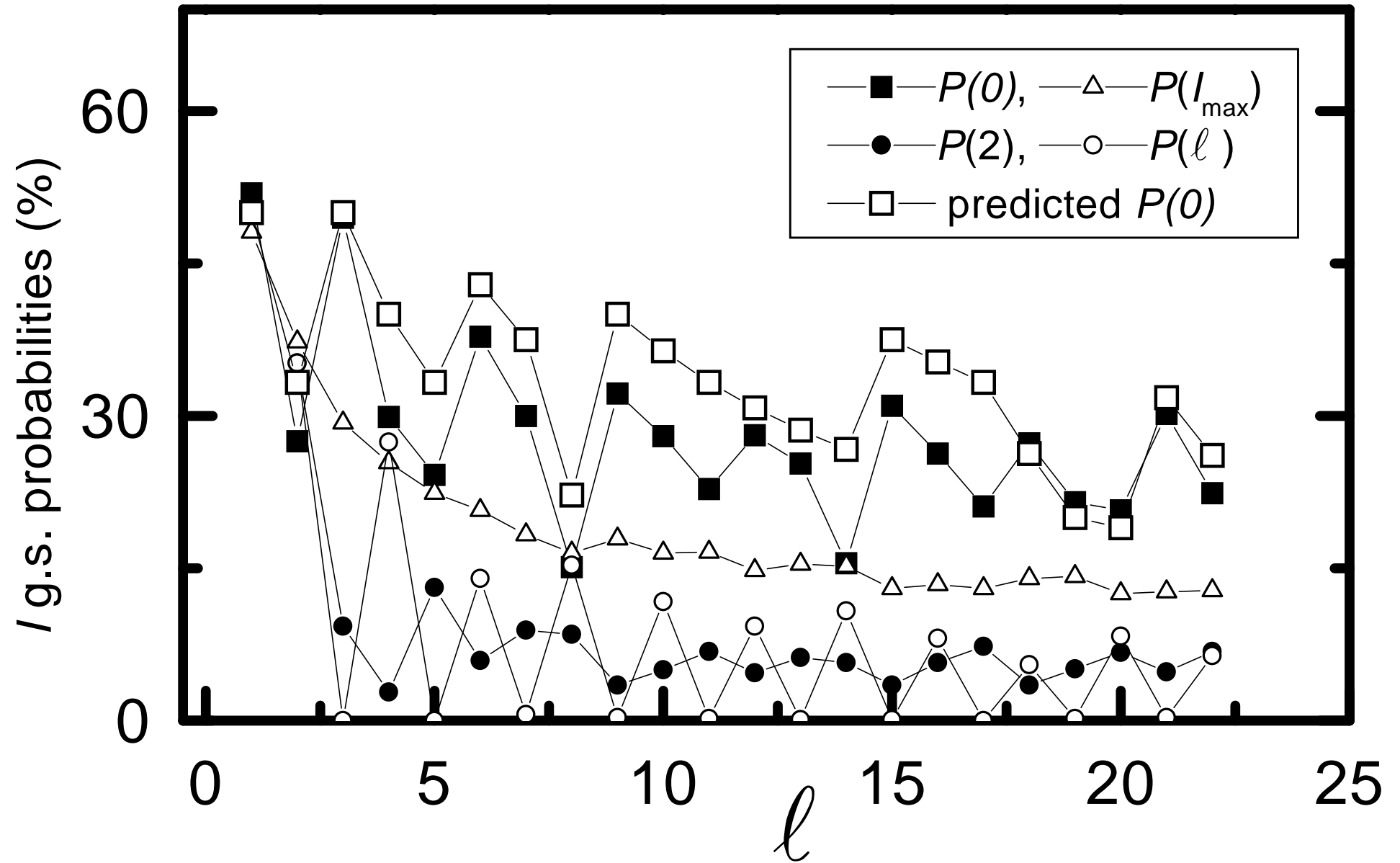
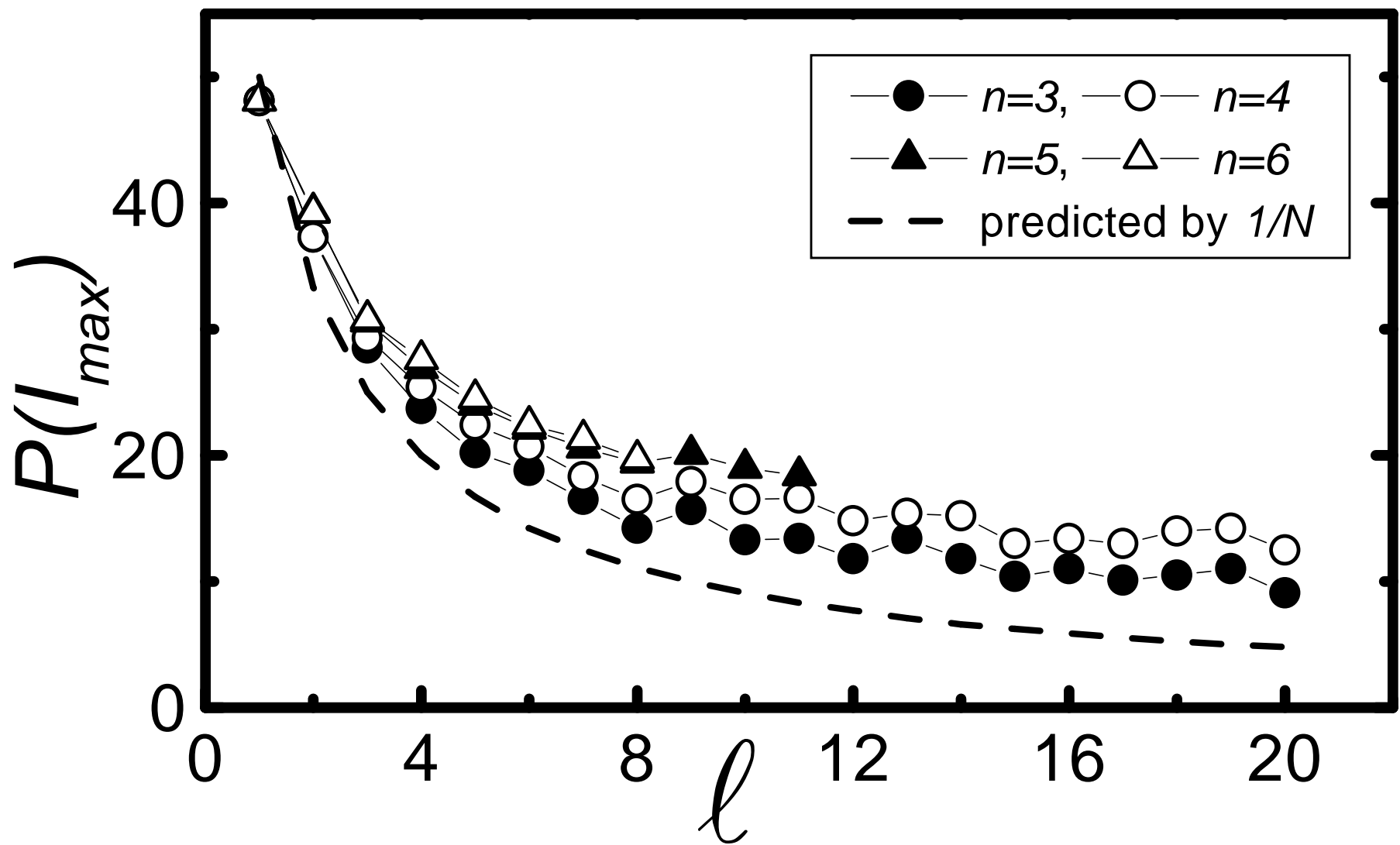
\end{document}